\documentclass[prd,twocolumn,showpacs]{revtex4}
\usepackage{graphicx}
\usepackage{latexsym}
\usepackage{amsmath}
\begin{document}
\title{Energy conditions and current acceleration of the universe}
\author{Yungui Gong}
\email{yungui_gong@baylor.edu}
\affiliation{College of Electronic Engineering, Chongqing
University of Posts and Telecommunications, Chongqing 400065,
China}
\affiliation{GCAP-CASPER, Physics Department, Baylor University,
Waco, TX 76798, USA}
\author{Anzhong Wang}
\email{anzhong_wang@baylor.edu}
\affiliation{GCAP-CASPER,
Physics Department, Baylor University, Waco, TX 76798, USA}
\begin{abstract}

The energy conditions provide a very promising model-independent study of the 
current acceleration of the universe. However, in order to connect these conditions 
with observations, one often needs first to integrate them, and then find
the corresponding constraints on some observational variables, 
such as the distance modulus.  Those integral 
forms can be misleading, and great caution is needed when one interprets them
physically. A typical example is that the transition point of the 
deceleration parameter $q(z)$ is  at about $z \simeq 0.76$ in the $\Lambda$CDM model. 
However, with the same model when we consider   the dimensionless Hubble 
parameter $E(z)$, which involves the integration of $q(z)$, 
we find that   $E(z)$ does not cross the line of  
$q(z) = 0$ before $z = 2$. Therefore, to get the correct result,
we  cannot use the latter to determine the transition point. 
With these in mind, we carefully study the constraints from the energy 
conditions, and find that, among other things, the current observational data 
indeed strongly indicate that our universe has ocne experienced an accelerating 
expansion phase between the epoch of galaxy formation and the present.

\end{abstract}
\maketitle

\section{Introduction}

Ever since the discovery of the accelerated expansion of the universe by the
supernova (SN) Ia observations \cite{agr98}, many efforts have been made
to understand the mechanism of this accelerated expansion. Although
different observations all pointed to the existence of dark energy  
\cite{riess,astier,riess06, wmap3,sdss}, the nature of it is still 
a mystery. For recent review of dark energy models, one may refer to
\cite{DE}.

Due to the lack of satisfactory dark energy models, many  model-independent 
methods were proposed to study the properties of dark energy and the geometry 
of the universe \cite{ClassA,gong07,ClassB,ClassC}. In particular,
in the reconstruction of the deceleration parameter $q(z)$, it was found
that the strongest evidence of acceleration happens at the
redshift $z\sim 0.2$ \cite{ClassA,gong07}. The sweet spot of
the equation of state parameter $w(z)$ was found to be
around the redshift $z\sim 0.2-0.5$ \cite{gong07,ClassB}.

Another very interesting and model-independent approach  is to consider 
the energy conditions \cite{HE73}. Since these conditions do not require a specific 
equation of state of the matter in the universe, they provide  very simple
and model-independent bounds on the behavior of the (total) energy density, pressure and
look-back time as a function of red shift. As a matter of fact, even before the discovery 
of the acceleration of the universe,  studies of these conditions  
already led Visser in 1997 to  conclude correctly that  { current observations suggest 
that the ``strong energy condition" (SEC) was violated sometime  between the 
epoch of galaxy formation and the present. This implies that no possible 
combination of ``normal" matter is capable of fitting the observational data}
\cite{visser}.  Santos {\em et al}   \cite{alcaniz} further investigated
these conditions and found that all the energy conditions seem to have been
violated in a recent past of the cosmic evolution. On the other hand, assuming 
that the universe is flat and contains only dark matter and dark energy, Sen and 
Scherrer studied the constraints of the weak energy condition (WEC) on 
the evolution of the Hubble parameter and the coordinate distance, and  obtained 
an upper bound on $\Omega_{m}$ \cite{aasen}. As the authors themselves pointed 
out, this bound is generic and independent of the nature of the dark 
energy. Lately, we also investigated these conditions, and applied them to 
the 192 essence supernova Ia data \cite{gong07b}. In particular, we showed that 
the universe had once experienced an accelerated expansion period. From the
degeneracy of the distance modulus at low redshift, we also argued that the
choice of $w_0$ for probing the property of dark energy is misleading.
One explicit example was used to support this argument.

\begin{figure}
\centering
\includegraphics[width=8cm]{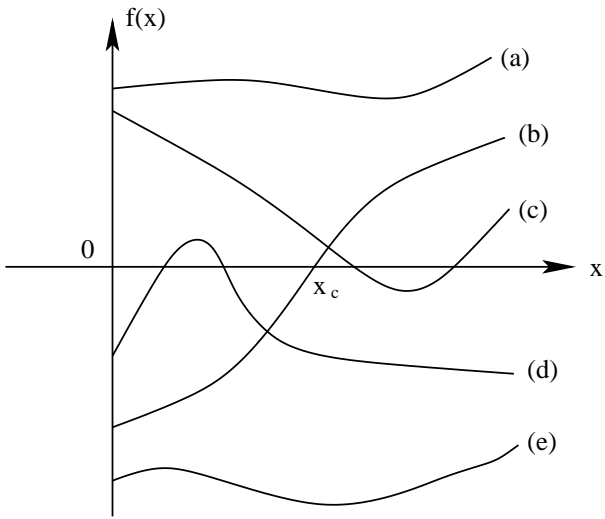}
\caption{The function $f(x)$ for several different cases. In Cases (a) and (c),
the integral $I(x) = \int_{0}^{x}{f(x')dx'}$ is always non-negative, while in Cases 
(d) and (e) it is always non-positive. In Case (b), $f(x)$ has a crossing point
at $x =x_{c}$,  but the crossing point of $I(x)$ is much great than $x_{c}$.}
\label{fig0}
\end{figure}

In this paper we would like to point out that, while such an approach is very promising,
one has to use these energy conditions with great caution. This is mainly because these 
conditions are local in terms of the expansion factor $a(t)$, and when we use them to 
study their constraints on some observational variables, such as the distance modulus
$\mu(z)$, we often need to consider their integral forms. Such integrated formulas can
be misleading, and result in wrong 
interpretations. To see this clearly, let us consider a function $f(x)$, which is 
smooth enough so that the integral $I(x) = \int_{0}^{x}{f(x')dx'}$ exists.  
Obviously, if $f(x) \ge 0$ for $x \in (0, x_{s})$, we must have $I(x) \ge 0$ 
for $x \in (0, x_{s})$ [Fig. 1(a)]. However, the inverse does not hold, that is the 
condition $I(x) \ge 0$ for $x \in (0, x_{s})$ does not imply $f(x) \ge 0$ 
for $x \in (0, x_{s})$. In particular, it does not exclude the possibility that $f(x)$ 
can be negative for some values of $x \in (0, x_{s})$. Case (c) in Fig. \ref{fig0}
shows explicitly this possibility. In fact, all what we can conclude from $I(x) \ge 0$  is 
that $f(x)$ must be non-negative for certain value(s) of $x \in (0, x_{s})$. 
Similarly, if $f(x) \le 0$ for $x \in (0, x_{s})$,  we must have $I(x) \le 0$ for 
$x \in (0, x_{s})$ [Fig. 1(e)], but the inverse is in general not true [Fig. 1(d)]. 

Another important remark is that the crossing points of $f(x)$ and $I(x)$ can be quite 
different. For example, $f(x)$ has a crossing point at $x_{c}$ along the curve (b) in Fig. 1, 
but clearly the crossing point of $I(x)$ must be much greater than $x_{c}$. 

With all of the above in mind, let us consider the constraints that the energy conditions
 impose. In particular, the paper is organized as follows. In section II, we consider
all the four energy conditions, and apply them first to the $\Lambda$CDM model and then
to a fiducial model. For the $\Lambda$CDM model, Fig. \ref{fig1} shows clearly that 
the deceleration parameter $q(z)$ passes the transition point at $z \simeq 0.76$. 
However, with the same model when we consider the dimensionless Hubble parameter 
$E(z)$, which involves the integration of $q(z)$, we find that $E(z)$ does not 
cross the line of $q(z) = 0$ before $z = 2$. Of course, the latter does not mean
that the transition must have happened at $z > 2$. Similar results can be obtained 
from our fiducial model given by Eq.(\ref{fiducial}), where $q(z)$ is negative during the 
period $0.1 < z < 0.15$. But, Fig. \ref{fig2} shows that $E(z)$ never crosses the line
of $q(z) = 0$. Applying our arguments to the 192 essence SN Ia data,
in section III we find that the data indeed strongly indicate that our universe 
has once experienced an accelerating expansion phase between the epoch of galaxy formation
and the present. In section IV we conclude the paper with some discussions.

\section{Energy Conditions}

The energy conditions  can be expressed as \cite{visser,alcaniz}

\begin{gather}
\label{nec}
{\rm NEC}\Leftrightarrow \rho+p\ge 0,\\
\label{wec}
{\rm WEC}\Leftrightarrow \rho\ge 0\ {\rm and}\ \rho+p\ge 0,\\
\label{sec}
{\rm SEC}\Leftrightarrow \rho+3p\ge 0\ {\rm and}\ \rho+p\ge 0,\\
\label{dec}
{\rm DEC}\Leftrightarrow \rho\ge 0\ {\rm and}\ \rho\pm p\ge 0.
\end{gather}
Combining with the FRW equation, for an expanding universe the SEC requires 
that
\begin{gather}
\label{sec1}
\rho+3p\ge 0 \Leftrightarrow q(t)=-\ddot{a}/(aH^2)\ge 0,\\
\label{nec1}
\rho+p\ge 0 \Leftrightarrow \dot{H}-\frac{k}{a^2}\le 0.
\end{gather}

The Hubble parameter $H(t)=\dot{a}/a$ and the deceleration
parameter $q(t)$ are related by,
\begin{equation}
\label{hubq}
H(z)=H_0\exp \left[\int^z_0 [1+q(u)]d\ln(1+u)\right],
\end{equation}
where the subscript $0$ means the current value of the variable.
Substituting Eq. (\ref{sec1}) into Eq. (\ref{hubq}), we find
\begin{equation}
\label{sech}
H(z)\ge H_0(1+z).
\end{equation}
On the other hand, the integration of Eq. (\ref{nec1}) yields
\begin{equation}
\label{sech1}
\quad H(z)\ge H_0\sqrt{1-\Omega_k+\Omega_k(1+z)^2},
\end{equation}
for redshift $ z=a_0/a - 1\ge 0$. For $z \ge 0$,
Eq. (\ref{sech}) implies Eq. (\ref{sech1}).
So we conclude that
\begin{gather}
\label{secres}
{\rm SEC}\Rightarrow H(z)\ge H_0(1+z),\\
\label{necres}
{\rm NEC}\Rightarrow H(z)\ge H_0\sqrt{1-\Omega_k+\Omega_k(1+z)^2}.
\end{gather}
However, the converses of Eqs. (\ref{secres}) and (\ref{necres}) are not true.
In particular, if Eq. (\ref{sech}) is satisfied, it does not mean that the SEC 
had never been violated, because Eq. (\ref{sech}) is the integration of 
Eq. (\ref{sec1}), similar to Case (c) illustrated in Fig. \ref{fig0}. 
In this case what we know is that  the SEC had once been satisfied. 
But, if the bound (\ref{sech}) is violated, then it is sure that the SEC 
had once been violated. By virtue of the same reasoning, the satisfaction 
of Eq. (\ref{sech1}) does not mean that  the NEC had {\em never} been violated, 
but does mean that the NEC had {\em once} been satisfied. Likewise, if this bound  
is violated, then  the NEC had {\em once} been violated.  

\begin{figure}
\centering
\includegraphics[width=8cm]{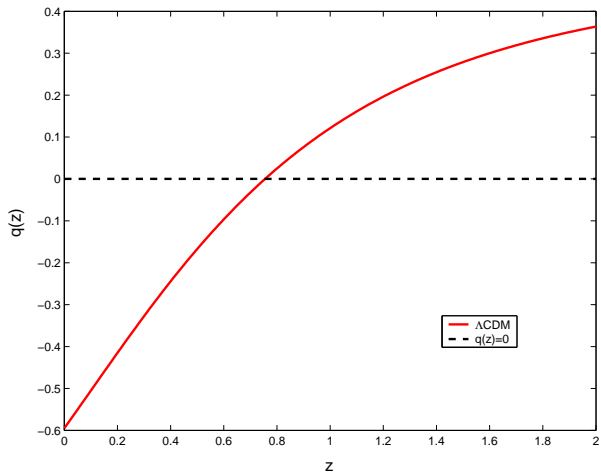}
\caption{The evolution of the deceleration parameter. The solid line
is for the flat $\Lambda$CDM model with $\Omega_m=0.27$ and the dashed line
is for $q(z)=0$.}
\label{fig1}
\end{figure}

\begin{figure}
\centering
\includegraphics[width=8cm]{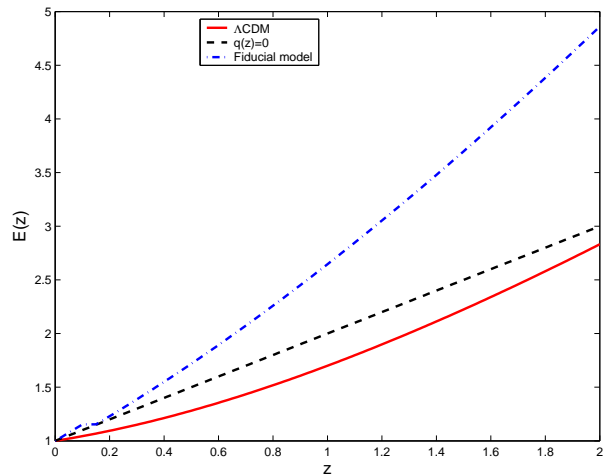}
\caption{The evolution of the dimensionless Hubble parameter $E(z)$. The solid line
is for the flat $\Lambda$CDM model with $\Omega_m=0.27$, the dashed line
is for $q(z)=0$, and the dash dotted line is for the Fiducial model (\ref{fiducial}).}
\label{fig2}
\end{figure}

\begin{figure}
\centering
\includegraphics[width=8cm]{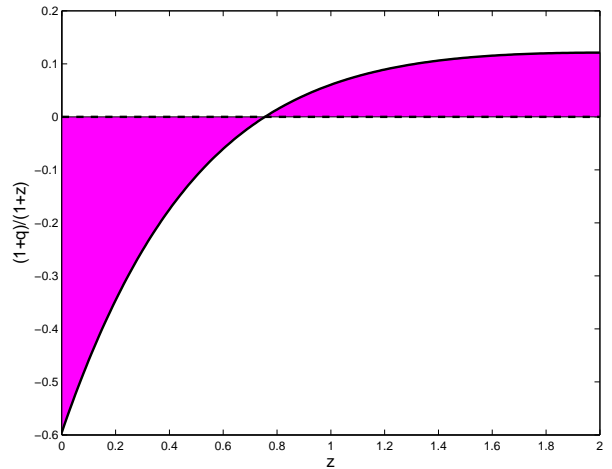}
\caption{The difference of the function $(1+q(z))/(1+z)$ between the flat $\Lambda$CDM model
and the model with $q(z)=0$.}
\label{fig3}
\end{figure}

These conclusions are very important, and we would like to use two specific
examples to help us further understand these key results. The first example
is the flat $\Lambda$CDM model with $\Omega_m=0.27$. The flat $\Lambda$CDM model
has accelerated expansion up to redshift $z=0.76$ and decelerated expansion for
$z> 0.76$. We plot the evolution of the deceleration parameter in Fig. \ref{fig1} 
where it clearly shows that $q(z)$ passes its transition point at
$z \simeq 0.76$. We also plot the evolution of the dimensionless Hubble 
parameter $E(z)=H(z)/H_0$ for the same model in Fig. \ref{fig2}. 
Even though when $z\ge 0.76$, $q(z)\ge 0$, we still have $H(z)\le H_0(1+z)$
up to $z\sim 2$ \cite{gong07b}. This may seem very strange, 
but it can be easily understood through
Fig. \ref{fig3}, where we plot the difference of the function $(1+q)/(1+z)$
between the $\Lambda$CDM model and the zero-acceleration model $q(z)=0$.
Because the Hubble parameter is related with the deceleration parameter
$q(z)$ by Eq. (\ref{hubq}), $H(z)$ is an integral of $q(z)$. Therefore, the
shaded area gives the value of $\ln(H_2/H_1)$, where $H_2$ denotes the Hubble
parameter of the $\Lambda$CDM model and $H_1$ denotes the Hubble parameter of the
model with $q(z)=0$.  The positive area of $2 \ge z \ge 0.76$ does not compensate 
the negative area of $z < 0.76$, so the total area is negative up to $z \sim 2$. 
This explains why $E(z)\le 1+z$ for the $\Lambda$CDM model even up to $z=2$.

The second example is the fiducial model
\begin{equation}
\label{fiducial}
q(z)=\begin{cases}
1/2, & z\le 0.1,\\
-1, & 0.1<z<0.15,\\
1/2, & z\ge 0.15.
\end{cases}
\end{equation}
Substituting this model into Eq. (\ref{hubq}), we obtain
\begin{equation}
E(z)=\begin{cases}
(1+z)^{3/2},& z\le 0.1,\\
1.1^{3/2}, & 0.1\le z\le 0.15,\\
[(1.1/1.15)(1+z)]^{3/2}, & z\ge 0.15.
\end{cases}
\end{equation}
The evolution of $E(z)$ for the fiducial model is shown in Fig. \ref{fig2}
by the dash dotted line. We see that even the bound (\ref{sech}) is satisfied 
for any given $z \ge 0$, $q(z)$ can still be negative in the interval $0.1 
\le z \le 0.15$. Thus, the bound (\ref{sech}) does not exclude the possibility 
that the universe had once experienced an accelerating expansion phase. From
this condition what we can really conclude is that the universe had once experienced a
decelerating expansion phase.

The $\Lambda$CDM model and the fiducial model (\ref{fiducial}) clearly show that
we must be very careful with the interpretation of the bounds (\ref{sech}) and (\ref{sech1})
derived from  the energy conditions. If the bound (\ref{sech}) is satisfied,
then we conclude that the SEC was once satisfied, although it is not necessarily 
always satisfied. The fiducial model (\ref{fiducial}) shows clearly that even if the 
bound (\ref{sech}) is satisfied, the SEC can still be violated during a certain
period of time.  If the bound (\ref{sech}) is violated,
what we are sure is that the SEC was once violated (but not necessarily always 
violated). The $\Lambda$CDM model shows that even if the bound (\ref{sech}) is violated, 
the SEC can still be satisfied for $z > 0.76$. 
Likewise, if the bound (\ref{sech1}) is satisfied, then we are confident that
NEC was once satisfied (but not necessarily always satisfied). 
If the bound (\ref{sech1}) is violated, then we are confident that
NEC was once violated.

\section{Cosmological applications of the Energy Conditions}

Now, let us consider the bounds on the luminosity distance. This  
was already discussed in \cite{gong07}. Here we would like to emphasize 
the key points derived in the last section.
We consider only the flat universe. Then, the luminosity distance is
given by
\begin{equation}
\label{lum}
d_L(z)=(1+z)\int_0^z\frac{dz'}{H(z')}.
\end{equation}
The extinction-corrected distance modulus is $\mu(z)=5\log_{10}[d_L(z)/{\rm Mpc}]+25$.
Substituting Eqs. (\ref{sech}) and (\ref{sech1}) into Eq. (\ref{lum}), we obtain
the upper bounds on the luminosity distance
\begin{equation}
\label{secbound}
H_0 d_L(z)\le z(1+z),
\end{equation}
\begin{equation}
\label{necbound}
H_0 d_L(z)\le (1+z)\ln (1+z).
\end{equation}
To compare these bounds with the 192 essence SN Ia data \cite{riess06}, we plot them in
the distance modulus redshift graph in Fig. \ref{fig4}. The region under the lower
solid line corresponds to the bound ({\ref{secbound}) and the region under the
upper solid line corresponds to the bound (\ref{necbound}). If all or some of the 
SN Ia data are inside the region under the lower solid line, it means that the 
universe had once experienced a decelerated expansion phase. If some or all of the 
SN Ia data are outside the region under the lower solid line, it means the universe 
had once accelerated. From Fig. \ref{fig4}, we see that some SN Ia are indeed 
outside the region under the lower solid line, so it is evident that the universe 
had once experienced an accelerated expansion. Note that due to the integration 
effect, even if some high $z$ SN Ia data are outside the region under the lower 
solid line, it does not mean that we have evidence of an accelerating expansion in 
the high $z$ region, as shown in Figs. \ref{fig1} and \ref{fig2}. 
Even if almost all the SN Ia data are outside the region under the lower solid line, 
it does not mean there is no evidence for past deceleration.

\begin{figure}
\centering
\includegraphics[width=8cm]{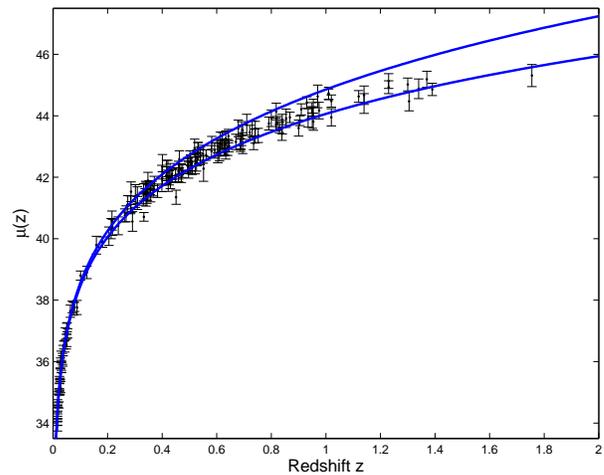}
\caption{The distance modulus $\mu(z)$. The solid lines correspond to the bounds 
from the SEC condition and NEC condition.}
\label{fig4}
\end{figure}

If all or some of the SN Ia data are inside
the region under the upper solid line, it means that the universe 
had once not experienced a super-accelerated expansion.
If some or all of the SN Ia data are outside the region under 
the upper solid line, it means the universe has once experienced a super-accelerated expansion.
Since the SN Ia data are in the region bounded by the two solid lines, we conclude that
{\em  the universe had once experienced an accelerating expansion phase,
and the acceleration is not always super-acceleration}. But this does not mean
that the universe has never experienced a period of  super-accelerated or 
decelerated expansion.

Now, let us turn to the bounds on the age of the universe derived from
 the energy conditions. The age of the universe is
\begin{equation}
\label{age}
t_0=\int_0^\infty\frac{dz}{(1+z)H(z)}.
\end{equation}
Substituting Eq. (\ref{sech}) into Eq. (\ref{age}), we get
\begin{equation}
\label{sect}
H_0 t_0\le 1.
\end{equation}
From the NEC condition for a flat universe, we get $H_0 t_0<\infty$. The current observational
values for $t_0$ and $H_0$ are $t_0=13.7^{+0.1}_{-0.2}$ and
$H_0=0.73^{+0.04}_{-0.03}\times (9.78{\rm Gyr})^{-1}$. Because $H_0^{-1}=13.4^{+0.6}_{-0.7}$,
so the current age of the universe is consistent with the bound ({\ref{sect}). However,
this does not mean that the current age of the universe is compatible with the SEC. 
The only conclusion   we can derive from this bound is that the SEC once held   
during the  past of the evolution of the universe.

If dark energy component satisfies the SEC, then we find
\begin{equation}
\label{desec}
E^2(z)=H^2(z)/H_0^{2}\ge \Omega_m (1+z)^3 +(1-\Omega_m)(1+z)^2,
\end{equation}
which results in
\begin{equation}
\label{omupper}
\Omega_m\le \frac{E^2(z)-(1+z)^2}{z(1+z)^2}.
\end{equation}
The results of $H(z)$ from  \cite{simon}, $H(1.53)=140\pm 14$, yield
the upper bound $\Omega_m\le -0.28\pm 0.08$. This upper bound is clearly
violated by current observations. Therefore, we conclude that SEC must
have once been violated. In other words, the universe had once experienced
an accelerated expansion. 

It is interesting to note that the WEC requires \cite{aasen}
\begin{equation}
\label{ssa}
\Omega_{m} \le \left.\frac{E^{2}(z) - 1}{(1+z)^{3} -1}\right|_{z = 1.53} 
= 0.18 \pm 0.05,
\end{equation}
which is also a little bit lower than that given by recent observations 
\cite{agr98,riess,astier,riess06,wmap3,sdss}.

\section{Discussion}

The energy conditions $\rho+3p\ge 0$ and $\rho+p\ge 0$ give lower bounds
(\ref{sech}) and (\ref{sech1}) on the Hubble parameter $H(z)$, and
upper bounds on the distance modulus $\mu(z)$. If some SN Ia data are outside
the region bounded by Eq. (\ref{sech}), then we conclude that the universe
had once experienced an accelerated expansion. If some SN Ia
data are outside the region bounded by Eq. (\ref{sech1}), then
we can tell that the universe had once experienced a super-accelerated 
expansion. In other words, the distance modulus-redshift
graph can be used to provide direct model-independent evidence
of accelerated and super-accelerated expansion.  
Therefore, the energy conditions provide direct
and model-independent evidence of the once-accelerated expansion phase. 
The bounds on the distance modulus also provide some directions for the 
future SN Ia observations. In particular, they can give some  
bounds on the age  of the universe and bounds on the distance 
modulus-red shift graph. 
 
Unfortunately, the method has also its own limitations.
For example, it does not provide us with any detailed information about 
the acceleration, nor the nature of dark energy. In addition,
Because the luminosity distance is an integral of the Hubble parameter,
the distance modulus does not give us  useful information about the exact
transition point of the universe from decelerated expansion to accelerated 
expansion.

\begin{acknowledgments}

We would like to thank J. Alcaniz for valuable discussions. YGG is supported 
by Baylor University, NNSFC under grant No. 10447008 and 10605042, SRF for ROCS, 
State Education Ministry and CMEC under grant No. KJ060502. AW is partially 
supported by VPR funds, Baylor University.

\end{acknowledgments}


\end{document}